\documentstyle[twocolumn,aps]{revtex}
\newcommand{\DD}{{\cal D}}
\begin{document}
\draft

\title{
The effect of forcing on the spatial structure and spectra of chaotically
advected passive scalars}
\author{Zolt\'an Neufeld, Peter H. Haynes and Guillemette Picard} 

\address{Centre for Atmospheric Science, 
Department of Applied Mathematics and Theoretical Physics,
University of Cambridge, Silver Street, Cambridge CB3 9EW, UK}

\date{\today}
\maketitle

\begin{abstract}
The stationary distribution of passive tracers chaotically advected by a
two-dimensional large-scale flow is investigated. The tracer field is forced
by resetting the value of the tracer in certain localised regions.  This
problem is mathematically equivalent to advection in open flows and results in
a fractal tracer structure.  The spectral exponent of the tracer field is
different from that for a passive tracer with the usual additive forcing (the
so called Batchelor spectrum) and is related to the fractal dimension of the
set of points that have never visited the forcing regions.  We illustrate this
behaviour by considering a time-periodic flow whose effect is equivalent to a
simple two-dimensional area-preserving map. We also show that similar
structure in 
the tracer field is found when the flow is aperiodic in time.  
%% This 
%%relationship is verified numerically in a simple model. 
%The effect of aperiodic
%fluctuations of the flow is also discussed. 
\end{abstract}

\section{Introduction}

It is well known that time-dependent flows with simple spatial
structure
produce complicated filamental patterns in advected passive
scalars through the phenomenon of chaotic advection \cite{chadvection}. 
Such patterns have been demonstrated in 
laboratory experiments \cite{exp1,exp2,exp3} and have also been simulated
numerically e.g. \cite{sim} (in the context of laboratory experiments)
and \cite{atmsim1,atmsim2} (in the context of the distribution of long-lived 
chemical species in the atmosphere).
Similar structures can be observed in satellite images
of oceanic plankton or sea surface temperature \cite{plankton}.
Although the two or three dimensional distribution of atmospheric
chemical species is not accessible to high-resolution observations, 
aircraft measurements
along one-dimensional transects  
give evidence of filaments and irregular structure with
strong spatial gradients \cite{atmsim1,Balluch,Tuck}.
The natural way to characterise these complex spatial distributions is by
investigating scaling properties of statistical quantities such as
Fourier power spectra or structure functions.

Using concepts from the field of chaotic dynamical systems has proved
to be useful in understanding the mechanism of mixing of passive
scalars (hereafter we shall use the term `tracer' to describe an
advected passive scalar)
in fluid flows with a smooth velocity field
and has led to the concept of chaotic advection \cite{chadvection} 
(Lagrangian chaos).
It has been recognised that the motion of fluid elements 
is typically chaotic even in very simple non-turbulent flows. 
Thus mixing appears as a result of the sensitivity of the particle
trajectories to their initial position. 
By stretching and folding fluid elements 
chaotic advection generates finer and finer structures
and tracer variance cascades down to small scales until it is dissipated
by molecular diffusion
and the system ends up in a perfectly mixed homogeneous state.

In order to maintain a nontrivial stationary state it is necessary
to apply some kind of forcing in the tracer evolution equation.
As we will show in this paper the specific form of this forcing  
has an important effect on the statistical properties of
the resulting tracer distribution.

The usual way to maintain tracer contrast (in numerical experiments and
theory) is by a large-scale {\em additive} forcing  
representing sources and sinks of the tracer. 
For this case, theory predicts a power spectrum $\Gamma(k) \sim k^{-1}$ 
for the tracer \cite{Batchelor} 
- the so called Batchelor spectrum  - 
in the viscous-convective range, that is between the
smallest characteristic length scale of the velocity field ($L$)
%%(in low Reynolds number viscous flows this is comparable to the
%%system size, or it is the so called Kolmogorov scale in turbulent flows)
and the characteristic length scale of diffusion ($l_D$).
Vulpiani has shown \cite{Vulpiani} that the Batchelor scaling is not 
restricted to the viscous-convective range of
turbulent flows but it is also valid for non-turbulent (e.g. time-periodic)
flows if the Lagrangian motion of the fluid elements is chaotic.
The $k^{-1}$ spectrum 
has been verified in recent numerical simulations 
\cite{Holzer,Antonsen,Yuan}. 
However, attempts to realise a $k^{-1}$ spectrum in laboratory experiments
have been inconclusive \cite{Miller,spectraexp}.

In a numerical study \cite{Pierre}, Pierrehumbert 
considered a different type of forcing - a {\em resetting} forcing -
in which, 
as fluid
parcels entered certain localised regions of the flow, 
tracer concentrations were reset to values associated with the relevant
region. 
This was motivated by giving examples of a number of experimental settings
where a {\em resetting} forcing seems to be more realistic than an 
{\em additive} one. Such examples are: $i.)$ thermal convection -
where the temperature of the fluid parcels is set to the temperature of the hot 
(lower) or cold (upper) boundary when they cross the thin 
thermal-boundary layers; $ii.)$ advection of dye concentrations -
where dye is introduced by diffusion from a solid source and
maintained at the saturation concentration in a diffusive boundary
layer. 
We note also that the 'resetting' forcing could 
be relevant 
for chemical tracers (e.g. ozone or nitrous oxide) in the
lower stratosphere, where polar and equatorial regions could play the
role of the forcing regions. 
%%, with strong chemical
%%activity in the polar and equatorial regions while they behave like
%%passive tracers elsewhere.

In \cite{Pierre} it was found that the power 
spectrum of the tracer was different from the expected 
$k^{-1}$, with spectral exponent between $-1$ and $-2$.
However, no precise theoretical explanation for the deviation 
from the Batchelor spectrum was given.

Tracer spectra steeper than $k^{-1}$ were observed in
recent experiments by Williams et al. \cite{spectraexp}.
The injection of tracer (dye) in these experiments also
seems to resemble the 'resetting' forcing.

In this article, we will show that the deviation from the Batchelor 
spectrum reported in \cite{Pierre}
is a direct consequence of the 'resetting' 
forcing.
Using a dynamicals system approach we show that this type of forcing
leads to the appearance of fractal patterns in the tracer 
distribution 
(similar to the ones characteristic of advection in open flows
\cite{open,open1}) 
that implies a spectral slope steeper than $-1$ (and dependent on
the fractal dimension of these patterns). 
In numerical simulations we model advection by a simple two-dimensional 
area-preserving map representing a time-periodic velocity field.
The effect of aperiodicity of the flow is also investigated, by 
considering a stochastic version of the map.

\section{The model}

The tracer concentration field $c({\bf r},t)$ in an 
incompressible flow is governed
by the equation:
\begin{equation}
{\partial c \over \partial t} + {\bf v}({\bf r},t) \cdot \nabla c =
%%S({\bf r}) + \alpha({\bf r}) (c - c_0({\bf r})) + 
{\cal F} + D \Delta c 
\label{advdiff}
\end{equation}
where ${\bf v}({\bf r},t)$ is the velocity field of the flow.
We assume
that ${\bf v}({\bf r},t)$ is smooth in space with only large scale 
structure (non-turbulent) and
time-dependent. Furthermore, the velocity field is assumed to
be independent of the tracer concentration (passive tracers).
For simplicity, we will consider two-dimensional flows 
but we expect that similar considerations apply to large-scale
flows in three dimensions.
Two-dimensional flows with large scale velocity structure 
are also relevant for
the stratosphere,
where the transport is mainly along two-dimensional isentropic
surfaces \cite{Pierre1,Pierre2}.

The term ${\cal F}$ on the right-hand side of (\ref{advdiff}) 
represents the forcing.
In case of 'additive' forcing ${\cal F} \equiv S({\bf r})$,
representing sources ($S >0$) and sinks ($S <0$) with a large
scale structure.
The 'resetting' forcing considered in this paper
can be represented mathematically as a limiting case of 
a relaxation to a concentration profile
$c_0({\bf r})$  
\begin{equation}
{\cal F} \equiv  \alpha({\bf r}) (c - c_0({\bf r}))
\label{forcing}
\end{equation}
where the rate of relaxation  $\alpha({\bf r})$ is
\begin{equation}
\alpha({\bf r}) = \left\{ \begin{array}{ll}
\infty & \mbox{if}\;\; {\bf r} \in \DD_i\\
0 & \mbox{otherwise}
\end{array} \right.
\label{alpha}
\end{equation}
(The case with constant relaxation rate ($\alpha({\bf r}) \equiv \alpha_0$)
representing some kind of homogeneous chemical or biological activity, was 
investigated in \cite{chemadv1,chemadv2}.)
Furthermore we assume, for simplicity, that $c_0({\bf r})$ is constant
in each forcing domain,
$c_0({\bf r}) \equiv C_i$ for ${\bf r} \in \DD_i$.
Thus the concentration in each fluid element is set to $C_i$ as
it enters the resetting domain $\DD_i$. 
%%and is conserved along its 
%%trajectory outside these domains. 
We will study the properties
of the concentration field outside the forcing domains.

We note, that tracer advection in a closed flow with
the resetting forcing applied is mathematically equivalent
to tracer advection in open flows with time-dependence
restricted to a finite mixing region, e.g.\ in the wake of a cylinder
\cite{open,open1}. For example, in \cite{open1} tracer concentrations
were imposed upstream to take one value in part of the inflow and
a second, different value on the remainder. The effect of advection
through the wake of the cylinder gave a complicated pattern of tracer
concentration downstream. 
In our case, the resetting domains play the role of the upstream
condition in the open flow problem, since fluid parcels emerge from
those domains with their value of tracer concentration specified. 
(The resetting domains also mimic the downstream escape in the open flow
problem, since when fluid parcels enter those domains their previous
value of tracer concentration is immediately forgotten 
due to the infinitely fast relaxation.)
%As the inflow region will have parts in different resetting domains $\DD_i$,
%the corresponding open flow has different
%values of the tracer in its inflow region. 
%Such situation was investigated in \cite{open1}.
Although in \cite{open1} the authors noted that the 
boundary between different values of tracer concentration 
had a fractal structure, the implication
for the tracer spectrum was not considered. 

The second term on the right hand side of (\ref{advdiff}) 
represents diffusion, which is assumed to be weak.
To make a more precise statement
let us denote by $\tau$ the characteristic time of the decay in time
of the number
of fluid particles that have not visited the 
forcing regions.  
We assume that the length scale of forcing ($L$) and the diffusive scale
($l_D \sim (D/\lambda)^{1/2}$) are well separated 
%%(high
%%Peclet number, $Pe? \equiv L/l_D$), 
so that
\begin{equation}
\tau \ll {1 \over \lambda} \log{L \over l_D}
\label{escaperesetting}
\end{equation}
where $\lambda$ is the rate at which chaotic advection generates
small scale structures, i.e. the absolute value of the
most negative Lyapunov exponent of the chaotic advection.
(In two-dimensional
incompressible flows this is equal to the positive Lyapunov
exponent, that is the rate of the exponential separation of nearby fluid 
elements \cite{chadvection}.)
The right hand side of (\ref{escaperesetting}) is the typical timescale needed 
for the large scale tracer 
variance to be dissipated by diffusion \cite{Pierre2}
that is much larger then the one needed for resetting. 
Thus, in this case, the fluctuations in tracer structure
is primarily affected by the resetting and not by diffusion.
Therefore we neglect diffusion in the followings 
($D =0$) but expect to obtain the same structure as it would be for 
nonzero diffusivity on scales
larger then $l_D$. 

In the absence of diffusion the concentration is conserved along
the trajectories of fluid elements (except for the resetting)
and it is convenient to study the problem (\ref{advdiff}) 
in a Lagrangian picture.
The concentration for a fluid particle depends only on which
resetting domain has last been visited by that particle.
Thus, in order to construct the asymptotic tracer distribution we
follow the backward trajectories by integrating
\begin{equation}
{{\rm d}{\bf r} \over {\rm d}t} = {\bf v}({\bf r},t)
\end{equation}
backwards in time from each point
until the trajectory enters one of the resetting domains and assign the
concentration corresponding to that domain to the point from where
the trajectory started.
The concentration distribution
along a line segment can be obtained in the same
way without having to calculate the whole two-dimensional field.
This allows us to reach high resolution easily for one-dimensional
transects.

%%To illustrate the characteristic features of the tracer field
%%forced by 'resetting' we first consider 
%%a time-periodic velocity field that is
%%a kinematic model of a meandering jet flow \cite{jet}. The resetting 
%%domains are taken to be parallel to the mean flow 
%%below and above the centre of
%%the jet, respectively.

Advection by a two-dimensional incompressible time-periodic velocity
field over one time period is equivalent to the action of a two-dimensional 
area-preserving map. 
%on the position of fluid particles. 
In general, the explicit form of this map cannot be obtained
analytically from the velocity field. 
For numerical convenience, in the detailed analysis of the tracer
field that follows we represent advection by a 
%%However one simple time-periodic
%%flow for which the explicit form of the corresponding map is available
%%is the piecewise steady flow given by periodic alternation of two
%%sinusoidal shear flows in the $x$ and $y$ direction, respectively.
%%The corresponding 
map ${\cal M}$ given by,
\begin{eqnarray}
x_{n+1} = x_{n} + a \sin{(2 \pi y_{n} + \phi)} \;\; {\rm mod} \;\; 1, 
\nonumber \\
y_{n+1} = y_{n} + a \sin{(2 \pi x_{n} + \phi)} \;\; {\rm mod} \;\; 1,
\label{map}
\end{eqnarray}
where $x_n$ and $y_n$ denotes the coordinates of a fluid
particle at discrete times $t=n T, \;  n=0,1,...$ and $T$ 
is the period of the flow. 
%%Since advection is represented by a map  
%%the position of fluid parcels is only defined at discrete
%%times separated by interval $T$. 
It is known that the orbits 
obtained by such iterations are typically chaotic.
We note, that the action of the
map (\ref{map}) is equivalent to 
advection for one period
by a piecewise steady flow given by periodic alternation of two
sinusoidal shear flows in the $x$ and $y$ direction, respectively.
%%Since we consider advection represented by a map  
%%the position of fluid parcels is only defined at discrete
%%times separated by interval $T$. 
%%For numerical convenience, in the detailed analysis  of the tracer
%%field that follows we represent 
%%Since advection is represented by the map ${\cal M}$ 
%%the position of fluid parcels is only defined at discrete
%%times separated by interval $T$. 

One characteristic of  
two-dimensional area-preserving maps is the existence of
quasiperiodic orbits that form KAM tori acting as transport barriers 
\cite{Ott}
separating different chaotic regions of the phase space.
%%Such transport barriers,
%%can be avoided by considering
%%$\phi$ as a random variable $\phi_n \in [0,2 \pi]$
%%taking on different values in each step of the iteration. 
In order to avoid such barriers  
we use a value of the parameter $a$ in (\ref{map}) 
that produces a single ergodic chaotic region
without visible KAM tori.
%%, while keeping
%%the flow periodic in time. 
An alternative approach (used in \cite{Pierre}) 
is to consider $\phi$ as a random 
variable $\phi_n \in [0,2 \pi]$
taking on different values in each step of the iteration.
We will first consider 
the deterministic case
($a = 0.6$, $\phi_n \equiv 0$),
and will then discuss the effect of the random parameter, 
representing aperiodic flows.

Since advection is represented by a map  
the position of fluid parcels is followed explictly only at discrete
times separated by interval $T$. 
%%The evolution of the tracer field according to the continuous time 
%%problem defined by  (\ref{advdiff}), (\ref{forcing}) and (\ref{alpha}) 
%%may be followed in the discrete time problem by resetting the value of
%%tracer concentration for fluid parcels moving under ${\cal M}$, i.e.\
%%at times $t=nT$, according to the rule that it is reset at time $nT$ to
%%the value corresponding to $\DD_i$ if $\DD_i$ was the last resetting
%%domain seen by the fluid parcel in the interval $(n-1)T < t \le nT$.
%%The latter condition is equivalent to the requirement that the fluid
%%parcel at time $nT$ lies  in a domain $\tilde \DD_i$, which contains
%%$\DD_i$ and which may be determined from knowledge of $\DD_i$ and of the
%%flow. 
%%One subtlety is that, although all choices of resetting domains $\DD_i$
%%for the continuous-time problem  have equivalent resetting domains
%%$\DD_i$ for the discrete-time problem, the reverse is not true. 
%%If it
%%is the continuous-time problem that is of interest, the resetting
%%domains $\DD_i$ should therefore, strictly speaking, not be chosen in
%%an ad hoc fashion. 
%%However in what follows it will become clear that
%%the precise geometry of the resetting domains is not particularly
%%important. 
We therefore follow Pierrehumbert (\cite{Pierre}) in resetting only at those
discrete times and also in 
considering resetting domains 
%%for the discrete-time problem 
of the form
of two parallel stripes of width $\epsilon$. 
%%(We also from here on drop
%%the distinction betweeen $\tilde \DD_i$ and $\DD_i$.)  
As we shall see later,
varying $\epsilon$, and hence the size of the domains, has a strong
effect on the resulting tracer distribution. [Note that every 
specification of the resetting domains in the time-continuous
problem defined by (\ref{advdiff}), with time-periodic flow, (\ref{forcing})
and (\ref{alpha}), corresponds some resetting domain in the time-discrete
problem. The converse, that every resetting domain in the time-discrete
problem corresponds to some resetting domain in the time-continuous problem,
applies provided the slight generalization is allowed that $\alpha$ is a
function of time as well space in the time-continuous problem.]

%%the size and structure of the resetting domain is an 
%%important parameter
%%for the tracer distrution.
%%in this problem. 
%%The resetting domains of the map (discrete in time), 
%%in fact,
%%represent the set of fluid particles that passed through
%%the resetting regions of the corresponding flow during the last period.
%%For simplicity we define the resetting domains directly for
%%the map but we note that for any structure of the resetting domains
%%of the time-continuous problem one can find the corresponding
%%domains for the discrete map. The inverse of this statement doesn't
%%hold, as may well be resetting domains of the map that have
%%no correspondent for the time-continuous problem (like the domains
%%with simple geometry that we considered, but since the exact shape
%%of these domains will not be important in the followings, we
%%prefer to take this simple choice).
 
\section{Results and discussion}

To illustrate the characteristic features of the tracer field
forced by 'resetting' we first consider
a time-periodic velocity field that is
a kinematic model of a meandering jet flow \cite{jet}. The resetting
domains are taken to be parallel to the mean flow
below and above the centre of
the jet, respectively.
In Fig.1 a snapshot of the concentration field
(obtained from the
backward iteration described in the previous section) is shown for
this meandering jet flow. The periodicity of the flow
implies a periodic tracer field for long times.
In the absence of barriers outside the forcing regions
almost all trajectories visit the
forcing regions in a finite time. Thus, the concentration field has only
two possible values ($C_1$ - 'black' and $C_2$ - 'white').
The tracer field is formed by elongated filaments
with a Cantor set
like fractal structure transverse to the filaments (see Fig.1b).
The width of the filaments varies 
in a broad range  of scales
showing that the forcing maintains 
an imperfectly mixed stationary state.
Similar structure can be observed
in the tracer field generated by the map (\ref{map}) (Fig. 2).

%%It is also visible in Fig.1. that the structure of the tracer fields 
%%is qualitatively the same in the two cases, 
%%but somewhat less regular (positions of the folds)
%%in the case of the random map (aperiodic flow). An obvious difference
%%in the fields is that while for the non-random map the tracer field
%%is invariant under the iterations of the map (after some transient time)
%%(or time-periodic in the time-continuous case) in the case of the random
%%map the tracer field is changing in time, but its qualitative appearance
%%remains the same.  

The boundary between the two regions of different concentration value
is in fact  the boundary between the basins of attraction of the  two
resetting domains for the time-reversed advection dynamics. The
structure of boundaries between basins of attraction of different
attractors is a well known problem in the field of chaotic dynamical
systems \cite{Ott,Grebogi,Bleher}. 
%%(both for dissipative [] and conservative case []).  
Complicated fractal structure of basin
boundaries appears as a consequence of non-attracting chaotic sets, 
usually associated with the phenomenon of transient chaos
\cite{transient}.

%%In our problem the advection dynamics is assumed to be chaotic. 
%%One characteristic of chaotic dynamics is the existence
%%of an infinite number 
%%of unstable periodic orbits \cite{?}
%%(whose number 
%%grows exponentially with the length of their period)
%%embedded in the chaotic region. 
%%The closure of the set formed by those periodic orbits
%%that have no intersections with the resetting zones
%%(i.e. those orbits together with the aperiodic orbits 
%%bouncing between
%%them) 
%%forms an invariant chaotic set

Let us consider the set ${\cal S}$ of points corresponding to orbits
that never visit the forcing domains
\begin{equation}
{\cal S} \equiv \{ {\bf r} | {\cal M}^n ({\bf r}) \not\in \cup_i \DD_i ,
\forall n = - \infty, \infty \},
\end{equation}
If the map ${\cal M}$ is the same in each iteration
the set ${\cal S}$ is, by definition, invariant under the iterations of
the map, ${\cal M}({\cal S}) = {\cal S}$.
Due to the assumed ergodicity of the advection, ${\cal S}$ is necessarily a
set of measure zero.
Such non-attracting invariant sets in chaotic systems 
(chaotic saddles) 
typically have a complex fractal geometry \cite{Grebogi,Bleher}.
In conclusion, the set of forcing domains selects 
a subset ${\cal S}$ of the existing chaotic trajectories that defines an
invariant fractal set.

Fig.3 shows the number of backwards iterations $n$ needed to reach
either of the resetting domains from points lying along a line segment.
The boundary between the regions with concentration $C_1$ and $C_2$ is
formed by points for which $n$ is infinite and coincides with the part
of the unstable manifold of ${\cal S}$ -- to be precise the union of
the parts of the unstable manifolds of each point in ${\cal S}$ that
lie between that point and the first resetting domain
encountered along the unstable manifold.  The results of estimating the
fractal dimension $D$ of the boundary along linear transects using the
so called box counting algorithm are shown in Fig.4.  
%%In principle the
%%boundary could be identified by considering the resetting time
%%distribution shown in Fig.3, but it is much more straightforward to
%%extract points on the boundary from the jumps in tracer distributions
%%along the linear transects (see Fig.2b). 

For a given flow, the geometry of the resetting regions - by selecting
an invariant chaotic set - determine the fractal dimension $D$.
The dependence of $D$ on the size of the resetting stripes $\epsilon$
is shown in Fig.5.
As the size of the resetting regions increases the fractal dimension
of the set corresponding to orbits avoiding these regions decreases.
If the resetting domains occupy the whole system, the distribution is
formed by a finite number of discontinuities ($D=0$).
(In fact, the fractal dimension $D$ may reach zero even if 
the forcing domains do not cover the whole chaotic region.)
In the limit of very small resetting domains ($\epsilon \to 0$)
the set of orbits avoiding them increases and ${\cal S}$
becomes space filling ($D \to 1$).

%%(this might be moved somewhere else)
%%We note, that advection in a closed flow with
%%the resetting forcing applied is mathematically equivalent
%%to the advection in open flows with time-dependence 
%%restricted to a finite mixing region.
%%In our case, as fluid particles enter the resetting domains
%%their identity is removed by the infinitely fast relaxation.
%%Thus, those regions in which particles enter and leave the resetting
%%domains can be seen as outflow and inflow regions of an open flow.
%%As the inflow region will have parts in different resetting domains,
%%the corresponding open flow will have inflow regions with different
%%values of the tracer. Such situation was investigated in \cite{}.
%%Although, the authors pointed out the fractal structure of
%%the boundary between tracers of different value its implication
%%to the spectra was not investigated.

The fractal dimension $D$ of the 
unstable manifold of the chaotic saddle can be related to dynamical properties
of the system by a variant 
of the so called Kaplan-Yorke formula \cite{kaplanyorke}
corresponding to non-attracting sets \cite{kaplanyorke1}
\begin{equation}
D = 1 - {\kappa \over \lambda},
\label{kaplanyorke}
\end{equation}
where $\lambda$ is the Lyapunov exponent on the chaotic 
saddle
and $\kappa$ is the escape rate, defined by
\begin{equation}
A(nT) = A_0 e^{-\kappa nT}
\end{equation}
where $A(nT)$ is the area occupied by points
that have not visited the resetting regions
by time $t=nT$,
($\kappa \equiv 1/\tau$). In our context $\kappa$ might more
appropriately be called the `resetting rate'. 
(More precisely, (\ref{kaplanyorke}) 
gives the so called information dimension 
\cite{kaplanyorke1}
$D_1$ of the invariant measure on the saddle, 
which can be somewhat smaller then the box counting dimension).
If the non-uniformity of the stretching is not too strong
the Lyapunov exponent corresponding to any subset of trajectories
is close to the Lyapunov exponent of the full advection dynamics.
In our case this means that $\lambda$ has only a weak dependence
on $\epsilon$ and the fractal dimension depends on the structure of the 
resetting
regions through the `resetting rate' $\kappa$.

In general, there is no simple formula for the resetting rate
since it depends on the detailed geometry of the resetting
domains and the structure of the chaotic trajectories.
In the time-discrete version of the problem, however, one can have a
situation where mixing is strong enough to redistribute the
non-reset particles almost uniformly between two resettings,
if $T \gg 1/\lambda$. Then the resetting rate can be approximated by
\begin{equation}
\kappa = - \ln{\left(1-\sum A_i\right)}, 
\label{escape}
\end{equation}
where $A_i$ is the area occupied by the forcing domain $i$
normalised by the area of the whole chaotic region.
This approximation seems to apply to our
numerical example ($\lambda = 2.75/ T$), as is shown
in Fig.5, where the fractal dimension 
of the boundary predicted by (\ref{kaplanyorke}, \ref{escape}) 
is compared with the
measurements.
%The condition above can also be satisfied in case of
%time-continuous advection if the resetting is only applied
%at a sequence of discrete times $nT$. In case of permanent
%resetting regions ($T=0$) the simple formula (\ref{escape}) 
%cannot hold as the condition $T \gg 1/\lambda$ is not satisfied.

The standard way to characterise such tracer distributions is
by their Fourier power spectra.
In our case the distribution along a one-dimensional transect is made up
by an infinite number of jumps between the two possible values, 
with the jumps located on a fractal set (see Fig.2b).
The power spectra of such structure has been discussed in 
\cite{Vassilicos}, and
it was found that there is a simple relationship between the
spectral slope ($\gamma$) and the fractal dimension ($D$) of the set
of the discontinuities
\begin{equation}
\gamma=2-D.
\label{spectra}
\end{equation}
A more detailed analysis of the relation between fractal dimension
and spectral slope in \cite{Pierre3} and in \cite{Namenson} 
(done in the context of advection by
compressible flows modelled by dissipative maps 
with fractal attractors) has shown that $D$ in (\ref{spectra})  
is the correlation dimension $D_2$ of the chaotic set.
Assuming further that multifractality 
\cite{multifractals} is weak ($D_2 \approx D_1$) and
using (\ref{kaplanyorke}) the spectral slope can be approximated as
\begin{equation}
\gamma \approx 1+{\kappa \over \lambda},
\end{equation}
where the second term represents the deviation from the
Batchelor scaling and is proportional with the
rate of 'resetting' $\kappa$.

For numerical convenience, instead of direct measurement of 
the power spectra we consider
the second order structure function \cite{Monin} defined by
\begin{equation}
S(r) \equiv \langle ( C(r_0+r)-C(r_0) )^2 \rangle
\sim (\delta r)^{\zeta}
\end{equation}
where $\langle .. \rangle$ denotes averaging over 
different values of the
coordinate $r_0$. The scaling exponent $\zeta$ is known 
to be related to the spectral slope 
by the simple relationship 
$\gamma = 1+\zeta$, if $1 < \gamma < 3$ \cite{Rose}.

Structure functions (calculated for different transects and their average) 
are shown in Fig.6.
Fluctuations of the local slope 
at large scales  
decay fast at smaller scales where structure functions calculated from
different transects tend to be parallel.
This can be explained by the fact that the statistics of 
the small scale structures 
is much better (as there are many along a transect) than that of the
few large scale features. 
In this sense, spatial averaging improves the scaling
at large or intermediate scales.

The scaling exponent $\zeta$ 
depends on the size of the forcing domains, taking values 
between $\zeta=1$, ($\gamma = 2$) and $\zeta=0$, ($\gamma = 1$).
Fig.7 shows the value of $\zeta$
as a function of the width, $\epsilon$, of the forcing domains.
In order to verify the relationship (\ref{spectra}) in our model
the measured value of the spectral slope (based 
on the structure function) is compared 
with the one predicted from
the fractal dimension measurement, for different
values of $\epsilon$ (Fig.7).
The relationship is confirmed with a very good accuracy.

We also performed numerical simulations with the stochastic version
of the map (\ref{map}) representing aperiodic flows.
The two-dimensional tracer field (Fig.8)
shows a pattern qualitatively similar to the one shown in Fig.2
but with a somewhat less regular character.
An obvious difference is that in this case the tracer field 
changes with time, although its
qualitative appearance remains the same.

In the case of aperiodic flows 
%%periodic orbits of the advected particles 
invariant chaotic sets cannot exist. Nevertheless, 
there may exist a set of chaotic orbits that never visit the resetting domains.
The set of points corresponding to these orbits 
now wanders from one iteration to the next,
but exhibits
the same type of fractal structure as seen in the 
non-random case for the set ${\cal S}$. (The resetting time, when
calculated for points lying along one-dimensional transect, also shows a
similar structure to that in Fig.3.)

Structure functions for the aperiodic case are shown in Fig.9.
%%The structure functions measured along different transects of the same field 
%%(at a fixed time $n T$)
%%are similar to the ones observed in the deterministic case (they tend to be
%%parallel at small scales, Fig 7a). 
As a consequence of aperiodicity, 
the spatially averaged structure function, measured at fixed time, is
now time-dependent. Furthermore, its local slope 
fluctuates even at relatively small scales. (Compare with Fig.6b.)
Nevertheless, the time-average of these structure
functions exhibits clear scaling (Fig.9b). 
(Note, that the slope of the temporally averaged structure function is
different from that of the non-random map.)
%%, although the scaling
%%properties of the non-random map are independent of the value of
%%a constant $phi$.)
Similar fluctuations of the spectrum 
have also been
observed in the case of attractors of dissipative random maps \cite{Namenson}
and in experiments with floating particles advected on the
surface of an aperiodic flow
\cite{Sommerer}. In both
cases, temporal averaging led to a good power law scaling.

The lack of good scaling for the spectra taken at fixed
times can be explained by the fact that scaling needs
a good sampling of the possible stretching histories
at each scale. 
Structures contributing to the
spectra at a certain scale $l$ have been advected
for a finite time $t_l \sim (1/\lambda) \log({L/l})$.
When the flow is aperiodic (i.e the map is different in
each time step) the time $t_l$ might not be enough 
for a good sampling of the possible velocity fields if $l$
is not small enough. 
This can be compensated by temporal
averaging, but obviously cannot be replaced by spatial averaging
in this case.
We conclude, that for the random map (or aperiodic flows)
the power law scaling could be restricted to rather small scales
but it is clearly present in the temporally averaged sense.
This also holds for the box-counting fractal dimension.
There is a well defined fractal dimension in a temporally averaged
sense
%%. This fractal dimension is 
that relates to the exponent of the
structure function as in the case of the non-random map.
%%(this was confirmed numerically (?)) 

\section{Summary and conclusions}

%Fractal structures have been previously observed in compressible
%flows, where tracers concentrate on a fractal attractor,
%and in open incompressible laminar flows, due to the existence of
%a non-attracting chaotic set (chaotic saddle).
%%Using results on fractal basin boundaries arising
%%in chaotic dynamical systems and ones relating fractal geometry
%%and power spectra
%%We have shown that 

We have shown that when resetting forcing is applied to a passive
tracer in a chaotic advection flow, the boundaries between different
tracer values have a fractal structure. Such structure is closely analogous to
that previously identified in open flows. 
As a consequence, when 'resetting' forcing is applied
(or, equivalently, when open flows are considered)
the spectra is steeper than the $k^{-1}$ predicted
by Batchelor for the case of additive tracer forcing. 
The deviation can be expressed in terms of characteristic exponents
of the advection and forcing dynamics.
The physical mechanism for the steeper spectra is that most of the tracer
variance is destroyed by the resetting forcing (or escape, in open flows)
before reaching the
dissipative scale. This effect is clearly visible in the tracer
distribution (Fig. 2) which contains large smooth regions,
while strong gradients (and hence diffusive dissipation rates) are
concentrated, in the limit of diffusivity tending to zero, 
on a fractal set of measure zero 

The implication of our results is that the form of the tracer forcing
can strongly affect the spectral slope of the advected tracer. 
Indeed, our results could explain why some
previous numerical experiments and most of the previous laboratory experiments
have failed to reproduce 
the $k^{-1}$ spectrum. Our numerical example shows that
this is definitely the case for 
the numerical results presented in \cite{Pierre}. 
The tracer forcing used in the experiment \cite{spectraexp}
also seems to be similar to our considerations. 
Although in \cite{spectraexp} the authors try to explain the unexpected 
deviation from the Batchelor spectra
as an intermittency effect, they recognise that the deviation
is much stronger then what one would expect for intermittency
corrections. This was confirmed recently in \cite{Yuan}, where
the effect of intermittency (non-uniform stretching)
on the spectra was investigated.
Non-space-filling, fractal,
scalar interface in the Batchelor regime
was also detected in jet flows \cite{Sreene} where the 
open
flow type behavior, equivalent to the resetting, should apply.
A jet flow was also considered in \cite{Miller} reporting 
spectra steeper then $k^{-1}$ in the Batchelor regime.
The experimental implementation of the purely additive forcing 
seems to be difficult from technical point of view and this
could be the reason why the Batchelor spectra was never unambiguously
observed in laboratory experiments. 

Of course, the resetting forcing is an extreme limit.
In reality, the relaxation to a certain tracer value
cannot be infinitely fast. We expect, however, that
similar structures would be observed whenever there
is a strong spatial inhomogeneity of the relaxation rate
$\alpha({\bf r})$, having values much smaller
than the Lyapunov exponent, $\lambda$, in some
mixing dominated regions and being much larger
then $\lambda$ in others where the relaxational forcing (resetting) would
have the main effect.

The 'resetting' model may also be a useful first approximation
for chemical tracers in the lower stratosphere. 
The observed spectra \cite{Bacmeister}, 
$\gamma \approx 1.8$, fits into the range
of our prediction ($1 < \gamma < 2$). Furthermore, 
cliff-like structures resembling the ones seen in our simple
model are characteristic of stratospheric aircraft data.
Nonetheless those strucutres are not separated by smooth regions.
The forcing mechanism for stratospheric chemical tracers is likely 
to be far more complex than that considered here and needs further
investigation.

{\bf Acknowledgements:} The Centre for Atmospheric Science is a joint
initiative of the Department of Chemistry and the Department of Applied
Mathematics and Theoretical Physics. ZN is supported by the UK Natural
Environment Research Council (grant: GR3-12531). GP visited Cambridge with
support from Ecole Polytechnique, Palaiseau. PH and ZN acknowledge the
support of the European Science Foundation in allowing them to attend a
Study Centre of the Programme on Transport in Atmospheres and Oceans. 
Further support for this work came from the EU (through the TOASTE-C
programme). 

%%Open flows, with time-dependence restricted to a
%%finite region (mixing region),
%%are known to produce fractal tracer structures as a
%%consequence of the chaotic repeller
%%existent in the mixing region of the
%%flow \cite{open}. 
%%We have shown
%%that when the resetting forcing is applied in closed flows
%%the same type of chaotic repeller and the corresponding
%%fractal features exist. The stationary tracer field is made
%%up by filaments separated by jumps sitting on a fractal set. 
%%On the other hand there is a theory for the spectral exponent
%%of such distribution. According to this the 
%%corresponding power spectra can
%%be anything between $k^{-1}$ and $k^{-2}$ and is, 
%%in general, different (steeper) from the Batchelor spectra
%%corresponding to the additive forcing.
%%This explains
%%the numerical results of Pierrehumbert.

% Figure 1a -- qq1.ps
% Figure 1b -- qq2.ps
% Figure 2a -- qq3.ps
% Figure 2b -- qq4.ps
% Figure 3  -- qq5.ps
% Figure 4  -- qq6.ps
% Figure 5  -- qq7.ps
% Figure 6  -- qq8.ps
% Figure 7  -- qq10.ps
% Figure 8  -- qqx.ps
% Figure 9a -- qq12.ps
% Figure 9b -- qq13.ps

\begin{figure}
\caption{
Stationary tracer field $(a)$ (and blow-up $(b)$) constructed from 
$400 \times 400$ particles by calculating backwards 
trajectories until they reach one of the resetting regions.
The jet model flow and the parameters used are described in 
[25]. The forcing domains are,
$y > 2.3$ - black and
$y < -2.3$ - white. Periodic boundary conditions are used in the 
$x$ coordinate.}
\end{figure}

\begin{figure}
\caption{
Stationary tracer distribution $(a)$(before resetting)
using $400 \times 400$ particles
corresponding to the map (\ref{map})
using $a=0.6$ and $\phi = 0$. $(b)$, One-dimensional transect
for $x=0.75$. The resetting domains are stripes parallel to
the $y$ axis centered at $x=0.25$-black and $x=0.75$
with width $\epsilon=0.2$
}
\end{figure}

\begin{figure}
\caption{
Number of iterations needed to reach one of the forcing domains vs. initial
coordinates calculated for 2000 points on
the line segment shown in Fig. 2b.
}
\end{figure}

\begin{figure}
\caption{
Box counting measurement of the set formed by the discontinuities
of the tracer field (calculated with high resolution
using 65536 points) measured along three different segments 
parallel to the $x$ axis $y = 0.1;0.4;0.7$. 
The slope of the power law fit (dashed line) is
$D =0.67$. 
}
\end{figure}

\begin{figure}
\caption{
Fractal dimension vs. the width of the forcing stripes $\epsilon$.
Dashed line shows the prediction based on (\ref{escape}) and (\ref{kaplanyorke}),
$1+ \ln{(1- 2 \epsilon)} / \lambda$, using the measured value of the
Lyapunov exponent, $\lambda \approx 2.75$.
}
\end{figure}

\begin{figure}
\caption{
Structure functions calculated for three different transects 
$y = 0.1;0.4;0.7$ $(a.)$ and 
structure function averaged over 10 equidistant transects
parallel to the $x$ axis for $\epsilon = 0.2$ $(b.)$. The power
low fit (dashed line) has a slope $\zeta = 0.325$.
}
\end{figure}

\begin{figure}
\caption{
The spectral slope
$\gamma$, vs. the width of the forcing domains, $\epsilon$, 
based on the measurement of the structure function
scaling exponent, $\zeta$, (diamonds), 
and values
predicted from the fractal dimension of the boundary (crosses).
The analytical prediction, $1-\ln{(1- 2 \epsilon)} / \lambda$, 
($\lambda=2.75$) is also shown (dashed line).
}
\end{figure}

\begin{figure}
\caption{
Tracer distribution (before the resetting)
calculated at $400 \times 400$ points
using to the stochastic version of the map (\ref{map})
with $a=0.6$ and $\phi_n \in [0,2\pi]$. 
The resetting domains are stripes parallel to
the $y$ axis centered at $x=0.25$-black and $x=0.75$
with width $\epsilon=0.2$
}
\end{figure}

\begin{figure}
\caption{
Structure functions for the random case.
$(a.)$ Spatially averaged structure functions
at 5 different iterations, each calculated by averaging
aver 10 different transects. ($\epsilon = 0.2$).
$(b.)$ Temporal average over spatially averaged structure
functions calculated from 10 consecutive iterations.
The slope of the power law fit is $\zeta = 0.445$.
}
\end{figure}


\begin{thebibliography}{xx}

\bibitem{chadvection} H. Aref, J. Fluid Mech.\ {\bf 143} (1984) 1;
J. M. Ottino, {\em The kinematics
of mixing: stretching, chaos
and transport}, (Cambridge University Press, Cambridge, 1989);
A. Crisanti, M. Falcioni, G. Paladin and A. Vulpiani,
Nuovo Cimento, {\bf 14}, 1 (1991)

\bibitem{exp1} J. Chaiken, R. Chevray, M. Tabor and Q. M. Tan, Proc.
Roy. Soc., {\bf A408}, 165 (1986)

\bibitem{exp2} W. L. Chain, H. Rising and J. M. Ottino, 
J. Fluid Mech., {\bf 170}, 355 (1986)

\bibitem{exp3} J. C. Sommerer, H. C. Ku and H. E. Gilreath,
Phys. Rev. Lett., {\bf 77}, 5055 (1996)

\bibitem{sim} J. M. Ottino, C. W. Leong, H. Rising and P. D. Swanson,
Nature, {\bf 333}, 419 (1988)

\bibitem{atmsim1} 
D.W. Waugh, R.A. Plumb, R.J. Atkinson, M.R. Schoeberl,
L.R. Lait, P.A. Newman, M. Loewenstein, D.W. Toohey, 
L.M. Avallone, C.R. Webster, R.D. May, 
J.\ Geophys.\ Res.,
{\bf 99} 1071 (1994).

\bibitem{atmsim2} 
R.T. Sutton, H. Maclean, R. Swinbank, A. O'Neill, F.W. Taylor, 
J.\ Atmos.\ Sci., {\bf 51}, 2995 (1994).

%S. Edouard, B. Legras, F. Lefevre and R. Eymard,
%Nature, {\bf 384}, 444 (1996)

\bibitem{plankton} K. L. Denman and A. E. Gargett, 
Ann. Rev. Fluid. Mech., {\bf 27}, 225 (1995)

\bibitem{Balluch} M. G. Balluch and P. H. Haynes, 
J. Geophys. Res., {\bf 102}, 23487 (1997)

\bibitem{Tuck} A. F. Tuck and S. J. Hovde, Fractal behavior of ozone
wind speed and temperature in the lower troposphere,
Geophys. Res. Lett. (to appear, 1999)

\bibitem{Batchelor} G. K. Batchelor, J. Fluid Mech.,
{\bf 5}, 113 (1959)

\bibitem{Vulpiani} A. Vulpiani, Physica {\bf D38}, 372 (1989)

\bibitem{Holzer} M. Holzer and E. D. Siggia, Phys. Fluids,
{\bf 6}, 1820 (1994)

\bibitem{Antonsen} T. M. Antonsen, Z. Fan, E. Ott and E. Garcia-Lopez,
Phys. Fluids, {\bf 8}, 3094 (1996)

\bibitem{Yuan} G.-C. Yuan, K. Nam, T. M. Antonsen, E. Ott and
P. N. Guzdar, Power spectrum of passive scalars in two-dimensional
chaotic flows, preprint (1999)

\bibitem{Miller} P. L. Miller and P. E. Dimotakis,
J. Fluid Mech., {\bf 308}, 129 (1996)

\bibitem{spectraexp} B. S. Williams, D. Marteau and J. P.
Gollub, Phys. Fluids, {\bf 9}, 2061 (1997)

\bibitem{Pierre} R. T. Pierrehumbert, Chaos Solitons Fract., {\bf 4},
1091 (1993); also in {\it Chaos Applied to Fluid Mixing}, edited by
H. Aref and M. S. El Naschie (Pergamon, 1995)

\bibitem{open} E. Zemniak, C. Jung and T. T\'el,
Physica {\bf D76}, 123 (1994); 

\bibitem{open1} \'A P\'entek, T. Toroczkai, T. T\'el,
C. Grebogi and J. A. Yorke; Phys. Rev., {\bf E51}, 4076, (1995)

\bibitem{Pierre1} R. T. Pierrehumbert, Phys. Fluids, {\bf A3},
1250 (1991); 

\bibitem{Pierre2} R. T. Pierrehumbert and H. Yang, J. Atmos. Sci.,
{\bf 50}, 2462 (1993)

\bibitem{chemadv1} Z. Neufeld, C. Lopez and P. Haynes, 
Phys. Rev. Lett., {\bf 82}, 2606 (1999); Z. Neufeld, C. Lopez,
E. Hernadez-Garcia and T. T\'el, The multifractal structure of
chaotically advected chemical fields, preprint (1999), chao-dyn/9907023

\bibitem{chemadv2} M. Chertkov, Phys. Fluids, {\bf 10}, 3017 (1998);
Phys. Fluids, {\bf 11}, 2257 (1999)

\bibitem{jet} A. S. Bower, J. Phys. Oceanogr., {\bf 21}, 173 (1991)

\bibitem{Ott} E. Ott, {\it Chaos in dynamical systems}, (Cambridge 
University Press, 1993)

\bibitem{Grebogi} C. Grebogi, E. Kosterlich, E. Ott and J. A. Yorke,
Physica {\bf 25D}, 347 (1987)

\bibitem{Bleher} S. Bleher, C. Grebogi, E. Ott and R. Brown,
Phys. Rev. {\bf A38}, 930 (1988)

\bibitem{transient} T. T\'el, 'Transient chaos', in {\it 
Directions in Chaos}, Vol. 3, edited by H. Bai-Lin 
(Word Scientific, Singapore, 1991)

\bibitem{kaplanyorke} J. D. Farmer, E. Ott and J. A. Yorke,
Physica {\bf D7}, 153 (1983); 

\bibitem{kaplanyorke1} G. H. Hsu, E. Ott and C. Grebogi,
Phys. Lett. {\bf A127}, 199 (1988)

\bibitem{Vassilicos} J. C. Vassilicos and J. C. R. Hunt,
Proc. R. Soc. Lond. A {\bf 435}, 505 (1991)

\bibitem{Pierre3} R. T. Pierrehumbert, in {\it Nonlinear Phenomena
in Atmospheres and Oceans}, edited by R. Pierrehumbert and G. Carnavale
(Springer, New York, 1992)

\bibitem{Namenson} T. M. Antonsen, A. Namenson, E. Ott and
J. C. Sommerer, Phys. Rev. Lett. {\bf 75}, 3438 (1995);
A. Namenson, T. M. Antonsen and E. Ott,
Phys. Fluids {\bf 8}, 2426 (1996)

\bibitem{multifractals} T. T\'el, Zeit. Naturforsch. {\bf 48}, 714 (1982)

\bibitem{Monin} A. S. Monin and A. M. Yaglom,
{\it Statistical Fluid Mechanics} Vol. 2 (MIT Press, Cambridge Mass., 1975)

\bibitem{Rose} H. A. Rose and P. L. Sulem, J. Physique {\bf 39}, 441 (1978)

\bibitem{Sommerer} J. C. Sommerer and E. Ott,
Science {\bf 259}, 335 (1993); J. C. Sommerer, Phys. Fluids {\bf 8}, 
2441 (1996) 

\bibitem{Sreene} K. R. Sreenivasan and R. R. Prasad,
Physica {\bf D38}, 322 (1989)

\bibitem{Bacmeister} J. T. Bacmeister, S. D. Eckerman, P. A.
Newman, L. Lait, K. R. Chan, M. Loewenstein, M. H. Profitt and
B. L. Gary, J. Geophys. Res. {\bf 101}, 9441 (1996)

\end{thebibliography}
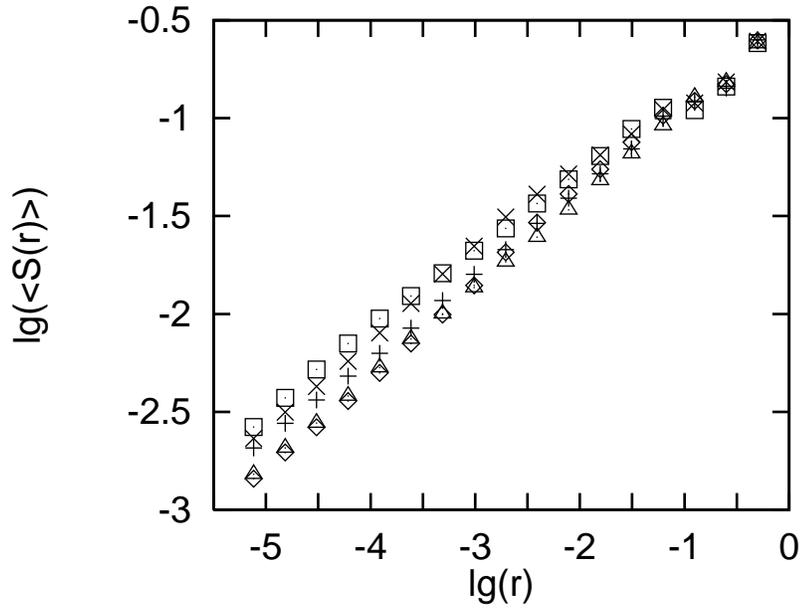
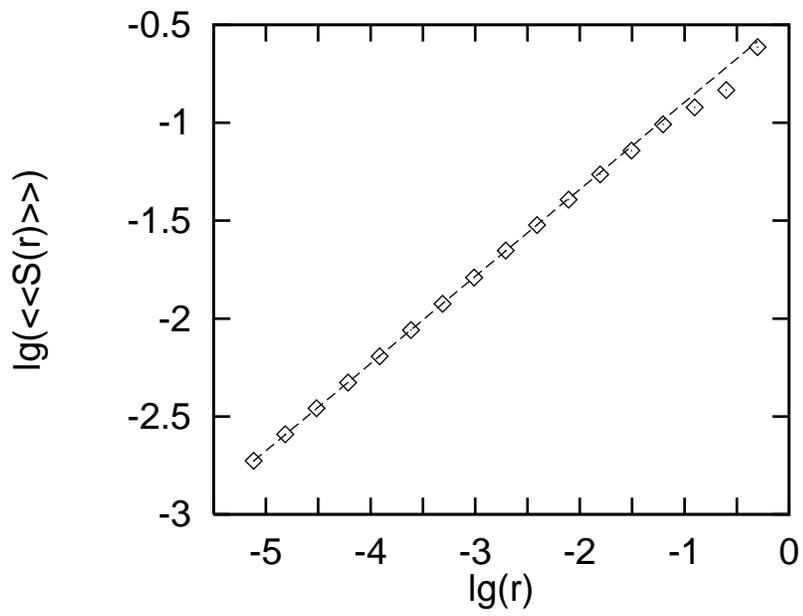
\end{document}